\documentclass[
letter,
twocolumn,
superscriptaddress,
amsmath,
amssymb,
prl,
nofootinbib
showkeys,
10pt,
floatfix
]{revtex4-2}

\usepackage{graphicx}
\usepackage{dcolumn}
\usepackage{bm}
\usepackage{xcolor}
\usepackage{ulem}
\usepackage{hyperref}
\usepackage{cleveref}
\usepackage{amsmath}
\usepackage{float}
\usepackage{wasysym}

\usepackage{svg}
\usepackage{amssymb}
\usepackage{scalerel}
\usepackage{caption}
\usepackage{setspace}
 \usepackage{multirow}
 \usepackage{booktabs}
 \usepackage{siunitx}
 \usepackage{physics}
 \usepackage{amsmath}
\usepackage{orcidlink}

\makeatletter
\pretocmd\frontmatter@keys@format{\addvspace{20\p@}}{}{}
\makeatother

\usepackage[font=small]{caption}

\begin{document}

\title{Self-hybridisation between interband transitions and Mie modes in dielectric nanoparticles}

\author{Christos Tserkezis\,\orcidlink{0000-0002-2075-9036}}
\affiliation{
 POLIMA---Center for Polariton-driven Light--Matter Interactions, University of Southern Denmark, Campusvej 55, DK-5230 Odense M, Denmark
}
\email{ct@mci.sdu.dk}

\author{P.~Elli Stamatopoulou\,\orcidlink{0000-0001-9121-911X}}
\affiliation{
 POLIMA---Center for Polariton-driven Light--Matter Interactions, University of Southern Denmark, Campusvej 55, DK-5230 Odense M, Denmark
}

\author{Christian~Wolff\,\orcidlink{0000-0002-5759-6779}}
\affiliation{
 POLIMA---Center for Polariton-driven Light--Matter Interactions, University of Southern Denmark, Campusvej 55, DK-5230 Odense M, Denmark
}

\author{N.~Asger~Mortensen\,\orcidlink{0000-0001-7936-6264}}
\affiliation{
 POLIMA---Center for Polariton-driven Light--Matter Interactions, University of Southern Denmark, Campusvej 55, DK-5230 Odense M, Denmark
}
\affiliation{
 Danish Institute for Advanced Study, University of Southern Denmark, Campusvej 55, DK-5230 Odense M, Denmark
}

\begin{abstract}
We discuss the possibility of self-hybridisation in high-index
dielectric nanoparticles, where Mie modes of electric or magnetic type can
couple to the interband transitions of the material, leading to spectral
anticrossings. Starting with an idealised system described by moderately
high constant permittivity with a narrow Lorentzian, in which self-hybridisation
is visible for both plane-wave and electron-beam excitation, we embark on a
quest for realistic systems where this effect should be visible. We explore
a variety of spherical particles made of traditional semiconductors such as
Si, GaAs, and GaP. With the effect hardly discernible, we identify two major
causes hindering observation of self-hybridisation: the very broad spectral
fingerprints of interband transitions in most candidate materials, and the
significant overlap between electric and magnetic Mie modes in nanospheres.
We thus depart from the spherical shape, and show that interband--Mie
hybridisation is indeed feasible in the example of GaAs cylinders, even
with a simple plane-wave source. This so-far unreported kind of polariton
has to be considered when interpreting experimental spectra of Mie-resonant
nanoparticles and assigning modal characters to specific features. On the
other hand, it has the potential to be useful for the characterisation of the
optical properties of dielectric materials, through control of the hybridisation
strength via nanoparticle size and shape.
\end{abstract}

\maketitle

\section{Introduction} 

Polaritonics is currently one of the most rapidly growing areas
in nanophotonics~\cite{ballarini_nanoph8,basov_nanoph10}, but also one
of the richest in physics~\cite{torma_rpp78,tserkezis_rpp83}. With
strong light--matter interactions lying at its heart, polaritonics
has drawn inspiration from quantum optics~\cite{Haroche_Oxford} to
evolve into an intriguing and fruitful combination of electrodynamics,
condensed-matter physics, and chemistry. Over the years, a plethora
of strong-coupling configurations has been introduced, ranging from
semiconductor quantum wells~\cite{bloch_apl73,khitrova_natphys2} and
quantum dots~\cite{stievater_prl87,santhosh_natcom7} coupled to mirror
or Bragg-reflector cavities, to excitons in organic 
molecules~\cite{pockrand_jcp77,lidzey_nat395,bellessa_prl93}
and two-dimensional (2D) materials~\cite{basov_sci354,goncalves_prb97,goncalves_aom8,
menabde_nanoph11} interacting with photonic crystals or plasmonic
nanostructures. The exciting possibilities that open by the interplay
between some material mode and confined light, and the subsequent
formation of half-light--half-matter states, is widely explored for
applications related to quantum and neuromorphic
computing~\cite{sanvitto_natmat15,opala_omex13}, molecular
photophysics~\cite{bhuyan_chemrev123} and polaritonic 
chemistry~\cite{hutchison_angchem51,yuen-zhou_jcp156}, control of
optical chirality~\cite{stamatopoulou_nscale14,baranov_acsphot10},
Bose--Einstein condensation~\cite{byrnes_natphys10,ramezani_josab36}
and polariton lasing~\cite{kenacohen_natphot4,arnardottir_prl125}.

Excitons are not the only matter components that can produce polaritons.
In principle, every polar excitation in a material can couple strongly
to the optical modes of a closed or even open cavity. Among the most
characteristic examples are phonon polaritons~\cite{foteinopoulou_nanoph8},
where phonons in the bulk or in van~der~Waals materials~\cite{dai_sci343,
hu_aom8} can couple either directly to light or to excitons in
molecules~\cite{dolado_natcom13}. Similarly, spins in microwave
cavities produce magnons~\cite{cao_prb91}, the collective excitation
of electrons in different subbands leads to intersubband
polaritons~\cite{todorov_prb85}, and Landau-level transitions in 2D electron
gases can couple to electric resonators and, under the influence of a strong
magnetic field, produce Landau polaritons~\cite{paravicini_natphys15}.
More complex polaritons involving the coupling of two quasiparticles
(one of the two usually being an exciton) include plexcitons in noble
metal nanoparticles (NPs) covered with an excitonic layer~\cite{fofang_nl8},
perovskite-exciton polaritons~\cite{su_natmat10}, and Mie-excitons in
high-index dielectric NPs~\cite{tserkezis_prb98,todisco_nanoph9},
possibly with a magneto-optical character~\cite{stamatopoulou_prb102}.

Recently, it was shown that an optical cavity is not strictly required
to achieve the modal coupling that produces polaritonic states. Indeed,
different optical modes in the same system can couple to each other,
just upon external excitation, e.g. with a laser~\cite{canales_jcp154}.
Such modes can be a combination of electronic and vibrational modes,
but self-hybridisation can also occur between Mie~\cite{Bohren_Wiley1983}
and vibrational modes in a water droplet~\cite{canales_2023}. A similar
effect in large dielectric spheres is related to excitons in
semiconductors interacting with the whispering-gallery modes of the
NP~\cite{platts_prb79}, or directly with Mie modes in
transition-metal-dichalcogenide (TMD) nanodisks~\cite{verre_natnano14}.
Inspired from these works, strong coupling was recently observed between
localised surface plasmons (LSPs) and interband transitions (IBTs) in
nickel (Ni) films and NPs~\cite{assadilayev_aom11}. In view of these
latest endeavours, it is reasonable to wonder why self-coupled polaritons
have never been reported in high-index dielectrics such as Si, where
spherical or cylindrical NPs support a richness of optical modes of
both electric and magnetic character~\cite{etxarri_oex19,evlyukhin_nl12},
and have been shown to be more flexible that plasmonic NPs in terms of
light--matter coupling~\cite{stamatopoulou_osac4}.

In this work, we show that interband--Mie coupling is indeed feasible,
and explain why its observation is experimentally challenging. We
begin with an ideal constant-permittivity sphere, where a superimposed
narrow Lorentzian mimics IBT resonances, and examine its optical
response with both plane-wave and electron-beam excitation. Having
shown, in this case, an anticrossing in both extinction and
cathodoluminescence (CL) photon-emission probability, and having thus
a clear fingerprint of strong coupling, we shift our attention to the
workhorse of Mie-resonant photonics, i.e. silicon (Si) nanospheres.
We identify two mechanisms that hinder experimental observation in
this case: the very large linewidth of IBT-related resonances in
the permittivity of bulk Si, and the fact that electric and magnetic
modes in spherical NPs have significant overlap. To overcome the
first issue, we explore other high-index semiconductors like gallium
arsenide (GaAs) and gallium phosphide (GaP). For the latter problem,
we break the spherical symmetry by designing dielectric pillars, in
which magnetic and electric modes can be well separated. Following
these actions, we finally manage to successfully identify the
self-hybridisation between IBTs and Mie modes. The feasibility of
this coupling introduces a new way for analyzing the intrinsic material
response of nanostructured materials. At the same time, it calls for
additional attention when explaining the origin of spectral features
and assigning modal characters to them, similarly to the emergence of
transition-radiation signals in CL that we reported 
recently~\cite{fiedler_nl22}.

\section{Ideal IBT--Mie hybridisation} 

\begin{figure*}[t]
\centering
\includegraphics[width=0.9\textwidth]{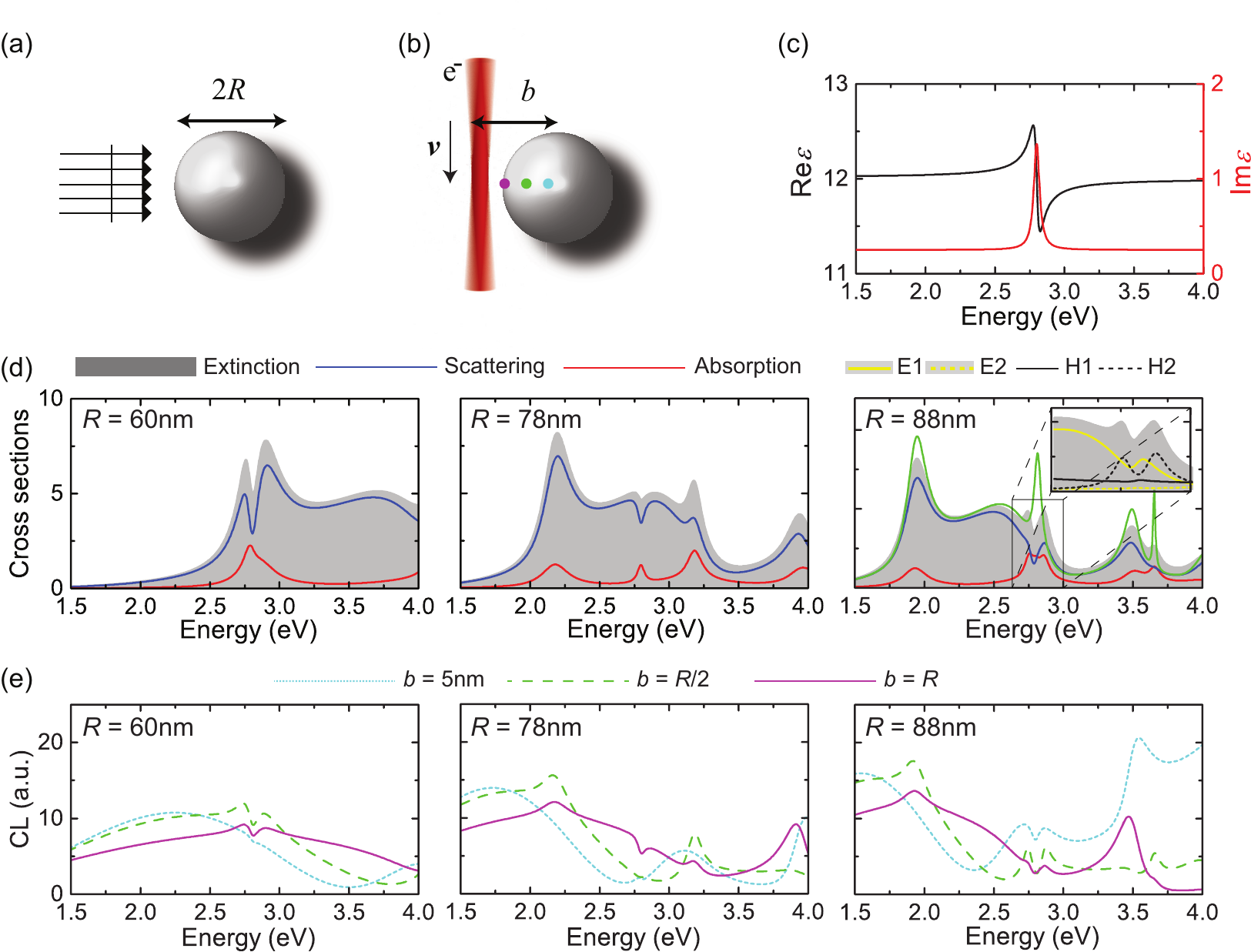}
\caption{
(a) A spherical NP of radius $R$ excited by a plane wave.
(b) The same NP as in (a), excited by an electron beam with velocity
$v = 0.33c$, passing near or through the NP at impact parameter $b$.
The coloured dots denote the three excitation spots corresponding to
the spectra shown in (e).
(c) Real (black line, left vertical axis) and imaginary (red line, right
axis) part of the permittivity of Eq.~(\ref{Eq:eps}).
(d) Normalised (to the geometrical cross section $\pi R^{2}$) extinction
(grey shaded areas), scattering (blue lines), and absorption (red lines)
cross sections, for spherical NPs described by the permittivity of (c),
for three different radii $R = 60$\,nm (left panel), $R = 78$\,nm
(middle panel), and $R=88$\,nm (right panel). In the rightmost panel,
the green line denotes the spectrum of an NP with the same radius but
in the absence of IBTs ($\varepsilon = \varepsilon_\infty=12$). The
inset zooms into the frequency window of the magnetic quadrupolar
mode (H2), and decomposes the extinction spectrum into contributions
from electric (E1 and E2) and magnetic (H1 and H2) dipoles and
quadrupoles (yellow solid and dashed, and black solid and dashed
lines, respectively).
(e) CL spectra (arbitrary units) for the same NPs as in (d), for 
three different impact parameters: grazing ($b = R$, purple solid lines),
traversing at $b = R/2$ (green dashed lines), or traversing near the
centre ($b = 5$\,nm, light-blue dotted lines) of the NP.
}
\label{fig1}
\end{figure*}

Let us begin by considering an ideal situation, of a nanosphere of radius
$R$ with a constant, moderately high relative permittivity $\varepsilon$
equal to $12$. Such an NP resembles Si in the infrared part of the 
spectrum~\cite{etxarri_oex19}, but more interesting physics takes place in
the visible and ultraviolet (UV)~\cite{green_semsc92}, including IBTs
that result in a negative permittivity in the UV and effectively
lead to a plasmonic behaviour~\cite{dong_nl19}. To mimic this, we add to
$\varepsilon$ a narrow Lorentzian term, so that the total, dispersive
permittivity becomes
\begin{align}\label{Eq:eps}
\varepsilon (\omega) = \varepsilon_\infty
- 
\frac{f \omega_{\mathrm{c}}^{2} }
{\omega^{2} - \omega_{\mathrm{c}}^{2} + 
\mathrm{i} \omega \gamma_{\mathrm{c}}}
.
\end{align}
In the above, $\omega$ is the angular frequency of the incident light,
$f = 0.02$ is the oscillator strength in the Lorentz model, $\omega_{\mathrm{c}}$
is the resonance frequency,  $\gamma_{\mathrm{c}}$ the damping rate, and
$\varepsilon_\infty = 12$. In order to have a reasonably broad resonance in
the frequency region where the first few Mie modes of NPs with radii varying
from $50$ to $100$\,nm appear, we choose $\hbar \omega_{\mathrm{c}} = 2.8$\,eV,
and $\hbar \gamma_{\mathrm{c}} = 0.05$\,eV. The real and imaginary part
of $\varepsilon$ described in Eq.~(\ref{Eq:eps}) are plotted in
Figure~\ref{fig1}c, corresponding indeed to a nearly constant permittivity
with a narrow resonance at $2.8$\,eV.

The nanosphere can be excited by either an incident plane wave, as
sketched in Figure~\ref{fig1}a, or by a swift electron beam,
passing near or through the NP with impact parameter $b$, as shown in 
Figure~\ref{fig1}b. In what follows, we assume that the electron beam
is modelled as a single electron travelling with speed $v = 0.33 c$
(with $c$ being the speed of light in vacuum); such a velocity corresponds
to a low acceleration voltage of $30$\,kV, which is typical in CL experiments.
The photon-emission probability for an electron emitted by the NP once it
has been excited by the electron beam is calculated here semi-analytically,
with the Mie-based method described in Ref.~\cite{stamatopoulou_2023}.

In Figure~\ref{fig1}d we scan over different NP radii, so as to gradually
match each of the first few Mie modes ---excited here with a plane wave---
to the energy of the Lorentzian. The first such mode is of magnetic dipolar
character (H1), originating from the emergence of displacement currents inside
the NP due to the excitation of bound charges in its bulk, with a phase 
difference as a result of retardation~\cite{evlyukhin_nl12}. For a radius
$R = 60$\,nm (left-hand panel in Figure~\ref{fig1}d), the H1 mode indeed
has its resonance at $2.8$\,eV; as a result of its interaction with IBTs,
a structure of two resonances with a dip in between them appears in the
extinction cross section spectrum (grey shaded area). A similar anticrossing
is present in the scattering cross section (blue line), which, due to the
nearly negligible losses in the bulk of the NP, dominates the extinction
spectrum. At the same time, the absorption cross section is characterised
by a broad resonance peaking at $2.78$\,eV, with a shoulder at $2.88$\,eV.
This feature in absorption suggests that indeed, there is interaction
between the two modes, albeit relatively weak; in principle, it would be
stronger if the NP were more lossy~\cite{tserkezis_prb98}.

Subsequently, in the middle panel of Figure~\ref{fig1}d we increase the
radius to $R = 78$\,nm, bringing the centre of the broad electric dipolar
(E1) Mie mode to the frequency of interest. What we see in this case is a
small dip in extinction and scattering, accompanied by just a single
resonance in  absorption, practically of the same absolute value as the
``hole'' left in the scattering spectrum. Clearly, there is no hybridisation 
between the two modes in this case, which is typical of what has been termed
enhanced or induced absorption~\cite{antosiewicz_acsphot1}. We have not
brought the two modes do interact, but rather we have added one mechanism
to subtract some of the energy corresponding to E1.

Finally, we increase the radius even further, to $R = 88$\,nm, for
which it is the magnetic quadrupolar mode (H2) of the NP that resonates
at $2.8$\,eV. In this case, the fingerprint of hybridisation is much 
better visible, since there is a double-peaked resonant feature in
both scattering and absorption spectra. In the right-hand panel of
Figure~\ref{fig1}d, we also include, with a green line, the extinction
spectra in the absence of any IBTs, i.e. when we set
$f = 0$ in Eq.~(\ref{Eq:eps}), to better demonstrate the two uncoupled
systems (H2 resonance in green line in Figure~\ref{fig1}d, and IBT
resonance in Figure~\ref{fig1}c). As further proof that it is indeed
the same mode, H2, that has split in two due to its interaction with 
the IBTs, the inset of the panel presents a decomposition
of the extinction spectrum into multipolar contributions, E1, E2, H1 and
H2 (E2 being the electric quadrupole). The E1 mode exhibits again the
tail of enhanced absorption discussed above (yellow solid line), but
H2 (black dashed line) is perfectly split into two hybrid resonances,
shifted by $0.05$\,eV on either side of the material resonance.

Since IBTs describe a loss mechanism directly related
to processes inside the bulk of the material, it is reasonable to ask
whether the hybridisation discussed above would be easier to observe
with a localised excitation, such as a focused electron beam passing
at a grazing trajectory, or even better penetrating inside the NP. In the
three panels of Figure~\ref{fig1}e we repeat the same radius-dependence
study as in Figure~\ref{fig1}d, but this time for CL calculations, for
three different impact parameters $b$: a grazing one (purple lines), a
beam travelling inside the NP at distance $b = R/2$ from the centre
(green lines), and one crossing the NP nearly at its centre (light-blue
lines) ---we set $b = 5$\,nm instead of $b = 0$\, for numerical reasons,
to ensure converged spectra. What we see in this case is that for both
the $R = 60$\,nm and the $R = 78$\,nm case, relatively narrow dips appear
at $2.8$\,eV, for both the $b = R/2$ and the $b = R$ electron trajectories.
But these are quite narrow to attribute to a hybridisation; the reason
for this weak interaction is rooted to the broadness of the spectral features
in CL, which is made worse for penetrating trajectories due to the
interference with transition radiation~\cite{fiedler_nl22}. This is most
clearly visible in the case of the $b \simeq 0$ trajectory, where
essentially only this periodic oscillation is visible. Nevertheless, for
the largest of the three NPs, $R = 88$\,nm in the right-hand panel of
Figure~\ref{fig1}e, it can still be seen that the H2 mode hybridises
with IBTs, and the strength of the interaction becomes larger for the
penetrating trajectory $b = R/2$. This suggests that CL spectroscopy
could be used as an alternative to plane-wave excitation in what follows,
where we shall explore realistic materials to establish if and when one
should anticipate to observe this behaviour in experiments. Nevertheless,
one should also keep in mind that, in our calculations, the electron
beam is modelled as a single electron travelling along a straight line,
free of any collisions. In realistic materials, one should always 
consider the possibility of the electron beam transferring momentum
and thus inducing indirect IBTs.

\section{Spherical particles of different materials} 

\subsection{Silicon}

\begin{figure}[h]
\centering
\includegraphics[width=\columnwidth]{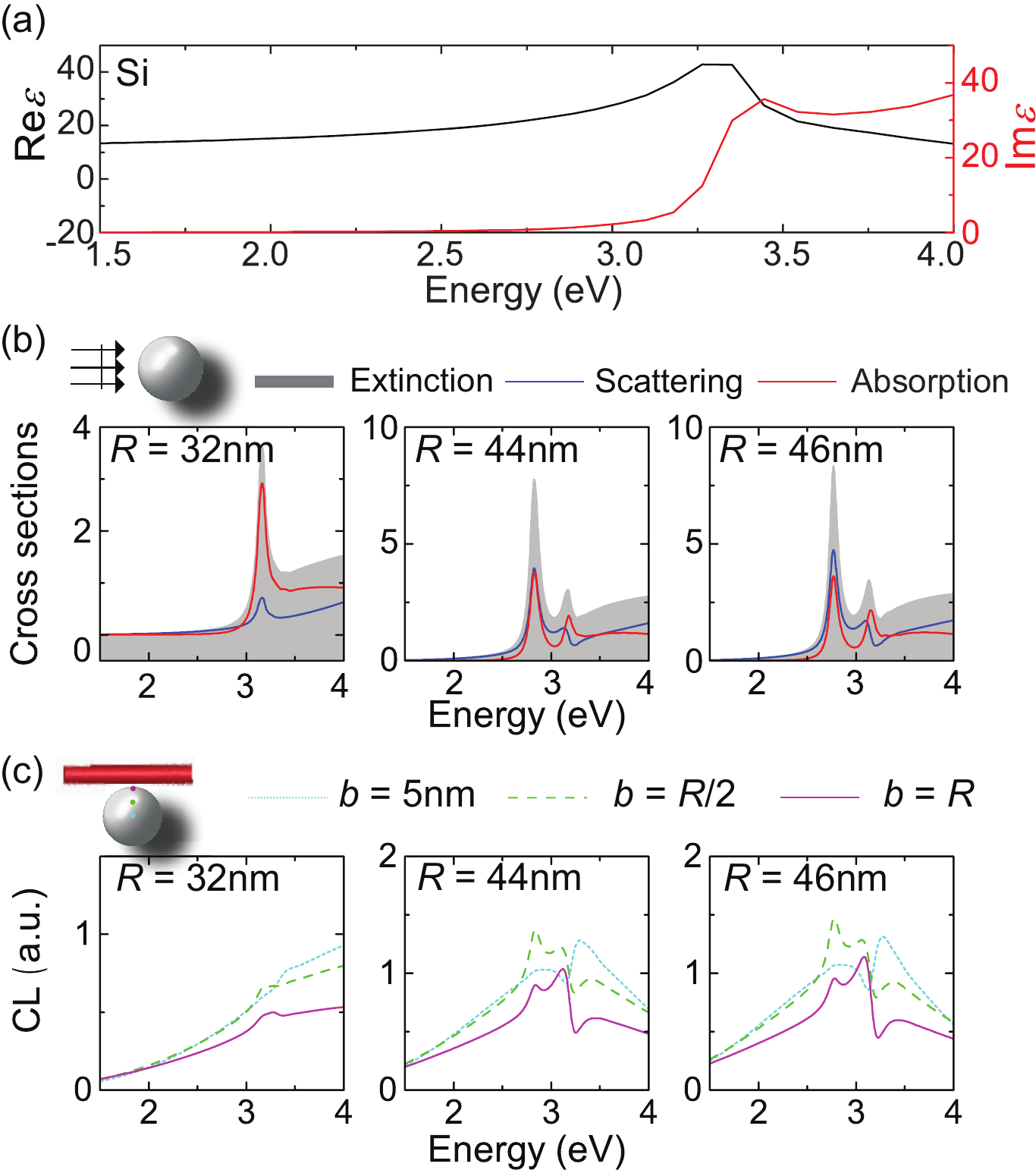}
\caption{
(a) Real (black line) and imaginary (red line) part of the relative
permittivity of bulk Si~\cite{green_semsc92}.
(b) Normalised extinction (grey shaded areas), scattering (blue lines),
and absorption (red lines) cross section of Si nanospheres of radii
$R = 32$, $44$, and $46$\,nm (left, middle, and right panel respectively),
upon plane-wave excitation.
(c) CL spectra (arbitrary units) for the same NPs as in (b), for 
three different impact parameters: grazing ($b = R$, purple solid lines),
traversing at $b = R/2$ (green dashed lines), or traversing near the
centre ($b = 5$\,nm, light-blue dotted lines) of the NP.
}
\label{fig2}
\end{figure}

Si is the dominant material in all theoretical and experimental activities
regarding Mie-resonant systems~\cite{liu_nanoph9,kivshar_nl22,dorodnyy_lpr17},
owing mostly to its availability in nature and its extensive use in modern
electronics. High-quality Si nanospheres with limited surface roughness
and good control over their dimensions are nowadays experimentally 
feasible~\cite{sugimoto_aom5}, enabling the detailed and accurate
characterisation of their optical properties. Nevertheless, IBT--Mie
resonance coupling has never been reported. The reason can be easily
understood by observing the permittivity of bulk silicon, shown in
Figure~\ref{fig2}a (reproducing data from Ref.~\cite{green_semsc92}).
The onset of IBTs appears at around $3.2$\,eV, but it is characterised
by a very broad, blunt resonance, and considerable losses (captured
by the imaginary part of the permittivity) all the way to the deep UV.
This already suggests that any geometrical resonance (such as the Mie
modes of interest here) will be severely damped, and only weakly couple
to the IBT resonance.

To make the situation even worse, the fact that IBTs appear at
such high energy means that small Si NPs are required in order
to align any Mie modes with them. This is indeed shown in the
far-field spectra of Figure~\ref{fig2}b, where we show that radii
as small as $\approx 30$\,nm are needed for the H1 mode to match
the energy of IBTs, and already at $R \leq 50$\,nm the ---hardly
distinguishable from E1--- H2 mode has also redshifted farther
than this energy. But for such small NPs, Mie resonances
(particularly the magnetic ones) are very weak: as we discussed
above, retardation plays a key role in the creation of the
displacement current which triggers these resonances, and
radii above $50-60$\,nm are desirable, as is also shown in
Figure~\ref{fig1}. It is then no real surprise that no
hybridisation can be observed in far-field spectra,
for none of the H1, E1, H2 modes (left, middle, and right
panel of Figure~\ref{fig2}b). But the same can be said for
electron-beam excitation and the corresponding CL spectra,
shown in Figure~\ref{fig2}c, again for three characteristic
electron-beam trajectories. Even though penetrating electron
trajectories could potentially excite the modes more efficiently
(as seems to be the case for the $b = R/2$ impact parameter,
green spectra in the figure), no evidence of modal interaction
can be observed---the sharp dip at about $3.2$\,eV is related
to the Kerker condition~\cite{shamkhi_prl122} and the interference
between the dipolar Mie modes of the NP.

\subsection{Other high-index semiconductors}

\begin{figure*}[ht]
\centering
\includegraphics[width=\textwidth]{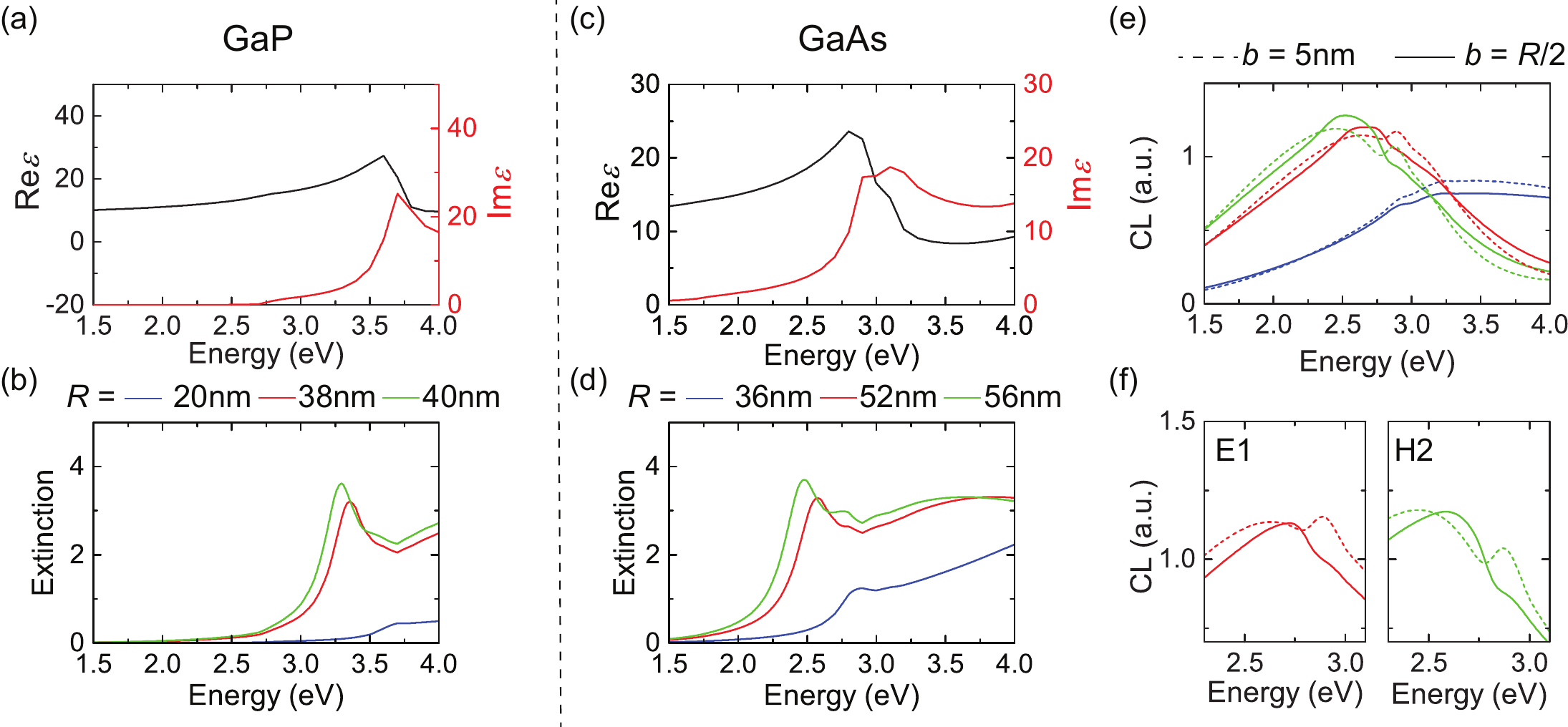}
\caption{
(a) Real (black line) and imaginary (red line) part of the relative
permittivity of bulk GaP~\cite{aspnes_prb27}.
(b) Normalised extinction cross section of GaP nanospheres of radii
$R = 20$, $38$, and $40$\,nm (blue, green, and red lines respectively).
(c) Real (black line) and imaginary (red line) part of the relative
permittivity of bulk GaAs~\cite{aspnes_prb27}.
(d) Normalised extinction cross section of GaAs nanospheres of radii
$R = 36$, $52$, and $56$\,nm (blue, green, and red lines respectively).
(e) CL spectra (arbitrary units) for the same NPs as in (d) (same
colour coding), for two different impact parameters: traversing at
$b = R/2$ (solid lines), or traversing near the centre ($b = 5$\,nm,
dashed lines) of the NP.
(f) Contribution to the spectra of the $R = 52$\,nm NP of panel (e)
from the electric-dipolar (E1) (left-hand panel) and the magnetic-quadrupolar
(H2) (right-hand panel) mode only.
}
\label{fig3}
\end{figure*}

Since we attributed the absence of any IBT--Mie hybridisation in
Si to the intrinsic properties of the material and, in particular,
the high (for our needs) energy at which IBTs appear and the large
linewidth of the resonance (if any resonance can be identified
above $3.5$\,eV), it is reasonable to explore other high-index
materials. To this end, we resort to Ref.~\cite{aspnes_prb27},
which is devoted to the measurement and characterisation of most
traditional semiconductors. Out of the materials encountered there,
we present in Figure~\ref{fig3} two cases: one that is not satisfactory
for our goal (GaP), and one that suggests that the desired hybridisation
could be visible (GaAs). Among the other possibilities, germanium (Ge)
has a resonance around $2.1$\,eV, albeit too broad for our needs,
while indium (In)-based compounds are characterised by more than one
well-defined resonance in a narrow energy window, and would not allow
for a clear interpretation of the calculated spectra. The GaP example,
depicted in Figure~\ref{fig3}, exhibits a relatively blunt resonance
around $3.5$\,eV, as can be seen in the permittivity of Figure~\ref{fig3}a.
Both of these characteristics make it a rather bad choice for our study
and, indeed, the extinction spectra of Figure~\ref{fig3}b do not suggest
any hybridisation with the (rather weak for these radii) Mie modes.

Unlike the aforementioned materials, GaAs appears to be a more
appropriate candidate in our quest. Its permittivity is dominated
by a relatively sharp resonance at lower energy, around $2.7$\,eV,
while a second resonance at slightly higher energy only appears as
a shoulder in the real part (Figure~\ref{fig3}c). The extinction
spectra of NPs with appropriate (larger than in the case of GaP)
radii ($R = 36$\,nm---blue line, $R = 52$\,nm---red line, and
$R = 56$\,nm---green line) suggest that there might indeed be
a possibility of interaction of all three modes, especially
E1 and H2, although it is not yet entirely clear whether the
resonance splitting indeed occurs, or it is mostly wishful thinking.
To further explore the situation, we calculate CL spectra, for two
different penetrating electron-beam trajectories ($b = R/2$---solid
lines, and $b = 5$\,nm---dashed lines), as shown in Figure~\ref{fig3}e
for all three NP sizes. Resonance splitting is now more visible
for the E1 and H2 modes, particularly when the electron beam travels
near the centre of the NP (in which case, the H1 mode is very weakly
excited). The two panels of Figure~\ref{fig3}f zoom into these
two splittings, showing only the contribution to the CL spectrum from
the corresponding term in the multipolar expansion of the field,
suggesting that it is indeed every individual mode coupling to the
IBTs. Nevertheless, this is  still a rather weak effect, which could
be masked in any experiment by fabrication imperfections and material
damping. What makes the interpretation of the spectra particularly
challenging, is the significant overlap between different multipoles.
To eliminate this source of ambiguity, we depart in what follows from
the spherical NP shape, and design cylindrical pillars with a moderate
aspect ratio, known to display well-separated magnetic and electric
Mie modes~\cite{staude_nn7}. 

\section{Separation of multipoles in cylinders}

\begin{figure*}[h]
\centering
\includegraphics[width=0.9\textwidth]{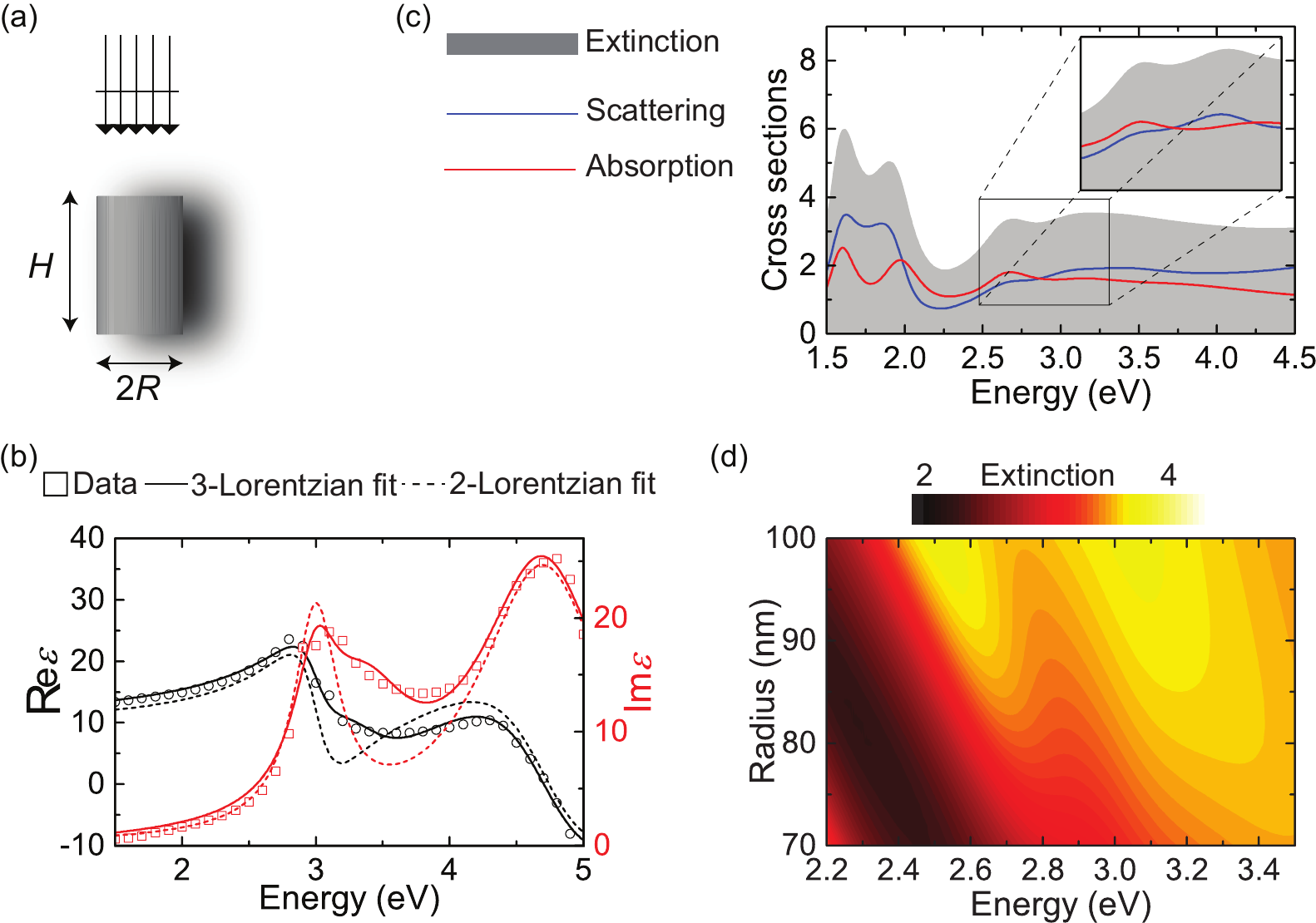}
\caption{
(a) Schematic of a GaAs cylindrical pillar with radius $R$ and height
$H$, illuminated by a plane wave coming from top.
(b) Real (black open circles) and imaginary (red open squares) part of
the permittivity of GaAs as obtained from experiment~~\cite{aspnes_prb27},
and Lorentzian fits to the data, with 3 (solid lines) or 2 (dashed lines)
Lorentzian oscillators.
(c) Extinction (grey shaded area), scattering (blue line) and absorption
(red line) cross section (normalised to the geometric cross section of
a sphere with the same volume) for a GaAs pillar as the one sketched
in (a), for $H = 200$\,nm and $R = 84$\,nm, described by the permittivity
given by solid lines in panel (b). The inset zooms in the
energy window where the E1 mode interacts with IBTs in GaAs, which
is now described by the permittivity shown with dashed lines in panel (b).
(d) Contour map of extinction of GaAs pillars with $H = 200$\,nm,
as a function of radius and energy, in the frequency region of IBTs.
}
\label{fig4}
\end{figure*}

In Fig.~\ref{fig4} we move from a spherical to a cylindrical shape.
We design cylindrical pillars of radius $R$ and height $H$, illuminated
with a plane wave from the top, as shown in the schematics of panel (a).
We focus on GaAs, as the most promising candidate according to the
discussion in the previous section. Instead of directly interpolating
the experimental data shown in Figure~\ref{fig3}c, we choose to fit
those data with  a multi-Lorentzian permittivity, similar to that of
Eq.~(\ref{Eq:eps}), with $\varepsilon_{\infty} = 2.169$ and three
oscillator terms with $f_{1} = 1.490$, $f_{2} = 2.379$, and
$f_{3} = 6.002$, resonant at $\hbar \omega_{1} = 3.007$\,eV,
$\hbar \omega_{2} = 3.398$\,eV, and $\hbar \omega_{3} = 4.731$\,eV;
the corresponding damping rates are $\hbar \gamma_{1} = 0.399$\,eV,
$\hbar \gamma_{2} = 0.836$\,eV, and $\hbar \gamma_{2} = 1.166$\,eV.
As can be seen in Figure~\ref{fig4}b, the fitting (solid lines) to
the experimental data is very good, and should reproduce accurately
all NP resonances. The reason for fitting the data is that it
conveniently allows us to selectively deactivate some of the features
at will. In particular, as we discuss next, it is particularly useful
to be able to disregard oscillator 2, and the corresponding resonance
expected at about $\hbar \omega_{2} = 3.398$\,eV; the corresponding
permittivity is shown in Figure~\ref{fig4}b with dashed lines, where,
in addition, we had to increase $f_{1}$ to $2.5$ to compensate for the
removal of the strong background that oscillator 2 adds to the
resonance of oscillator 1. For the calculation of the spectra,
traditional Mie theory is obviously no longer applicable; we thus
employ the nearest equivalent, namely the extended boundary condition
method (EBCM)~\cite{Mishchenko_Cambridge2002}, which is still based
on spherical-wave expansions. The simulation set-up and all convergence
parameters are as described in Ref.~\cite{todisco_nanoph9}. We should
note here that, within EBCM, the characterisation of modes in terms of
multipoles is straightforward, because the matrix that connects the
scattered to the incident field, though not diagonal, is almost
always dominated by a single element with specific angular-momentum
indices $\ell$ and $m$; alternatively, one can use Cartesian
multipoles~\cite{evlyukhin_prb94}.

In Figure~\ref{fig4}c we plot the extinction (grey shaded area),
scattering (blue line), and absorption (red line) spectra of a
pillar with $R = 84$\,nm and $H = 200$\,nm, a height that,
according to our calculations, is long enough to adequately
separate the different dipolar modes. Indeed, the E1 mode
appears now at about $2.8-3.0$\,eV, while H1 appears at much
lower energy, between $1.5-2.0$\,eV; the cylindrical shape has
in fact lifted the degeneracy of the mode in terms of magnetic
angular momentum $m$ number, and we now observe two separate
resonances. But, more importantly, the E1 mode seems to indeed
interact with IBTs at about $3.0$\,eV, and it produces a resonance
splitting, in both scattering and absorption, suggesting a real
hybridisation. Exploiting the spherical-wave decomposition that
is intrinsic in EBCM, we can confirm that the two resonances
come from the same $\ell$ and $m$ angular-momentum elements.
Another factor that need be considered, is the possibility
of the two resonances being related to the two different
resonances in the permittivity of GaAs, at frequencies
$\omega_{1}$ and $\omega_{2}$. This is where our fitting to
the experimental data proves important, since we can just
turn the resonance at $\omega_{2}$ off (by setting $f_{2} = 0$),
and explore the same system anew. The corresponding spectra,
just in the energy window of interest, are shown in the inset
of Figure~\ref{fig4}c, and indeed show the same double-resonant
behaviour. Finally, Figure~\ref{fig4}d shows a contour plot of
extinction versus nanodisk radius (always for the same height), and
how the two hybrid resonances emerge as we change the radius
and tune the E1 mode across the energy of IBTs.

It appears, therefore, that observation of hybridisation between Mie
modes and IBTs should be possible, at least in select conventional
materials, whose IBTs lie at energies for which Mie resonances
have fully developed, and in NP geometries that reduce the overlap
of these resonances. Such observations should be feasible both
in scattering and absorption measurements, but also in electron-beam
spectroscopies, since the crystal momentum is typically a few orders
of magnitude smaller that that of the electrons~\cite{Ashcroft_Harcourt1976},
and only direct transitions would be enabled. On the other hand, 
the emergence of this hybridisation could be exploited in the
characterisation of novel materials, where resonance splitting,
varying with NP size and shape, could provide information about the
strength and linewidth of resonances related to IBTs.

\section{Conclusion}

In summary, inspired from recent activities in self-hybridised
polaritons, we posed the question why no hybridisation between
Mie modes in high-index dielectric NPs and their intrinsic
IBTs has ever been reported. By considering an ideal, lossless,
constant-$\varepsilon$ sphere, we realised that this hybridisation
should indeed be feasible. We identified two major factors that
hinder its experimental observation: the very broad resonances
that characterise IBTs in most traditional high-index dielectrics,
and the significant overlap of Mie modes in spherical NPs. To
deal with the former, we considered a large variety of semiconductors,
and identified GaAs as the most suitable candidate. For the latter,
we departed from the spherical shape, and showed a clear hybridisation
in GaAs cylindrical NPs, with the effect in question being better
observable in the interaction of the E1 mode with IBTs. Recognition
of this possibility introduces another element of which one should
be aware when characterising NPs in terms of their modes.

\section*{Acknowledgement}
We are grateful for the submission deadline that made
the writing of this manuscript possible.

\section*{Funding}
N.~A.~M. is a VILLUM Investigator supported by 
\href{http://dx.doi.org/10.13039/100008398}{VILLUM FONDEN}
(Grant No.~16498).
The Center for Polariton-driven Light--Matter Interactions (POLIMA) is funded by the
\href{http://dx.doi.org/10.13039/501100001732}{Danish National Research Foundation}
(Project No.~DNRF165).

\section*{Author contributions}
C.~T. conceived the idea. C.~T., P.~E.~S., and C.~W. performed analytic
and numerical calculations. All authors contributed to analysing the data
and writing the manuscript. All authors have accepted responsibility for
the entire content of this manuscript and approved its submission.

\end{document}